\documentclass{moriond}

\def\Journal#1#2#3#4{{#1} {\bf #2}, #3 (#4)}

\def\PRL{\em Phys. Rev. Lett.}
\def\PRD{{\em Phys. Rev.} D}

\def\PNAS{{\em Proc. Natl. Acad. Sci. U.S.A.}}
\def\ApJL{{\em The Astrophysical Journal Letters}}
\def\ApJ{{\em The Astrophysical Journal}}
\def\AandA{{\em Astronomy and Astrophysics}}
\def\JHEA{{\em Journal of High Energy Astrophysics}}

\def\be{\begin{equation}}
\def\ee{\end{equation}}
\def\bea{\begin{eqnarray}}
\def\eea{\end{eqnarray}}

\newcommand{\Photo}{\includegraphics[height=35mm]{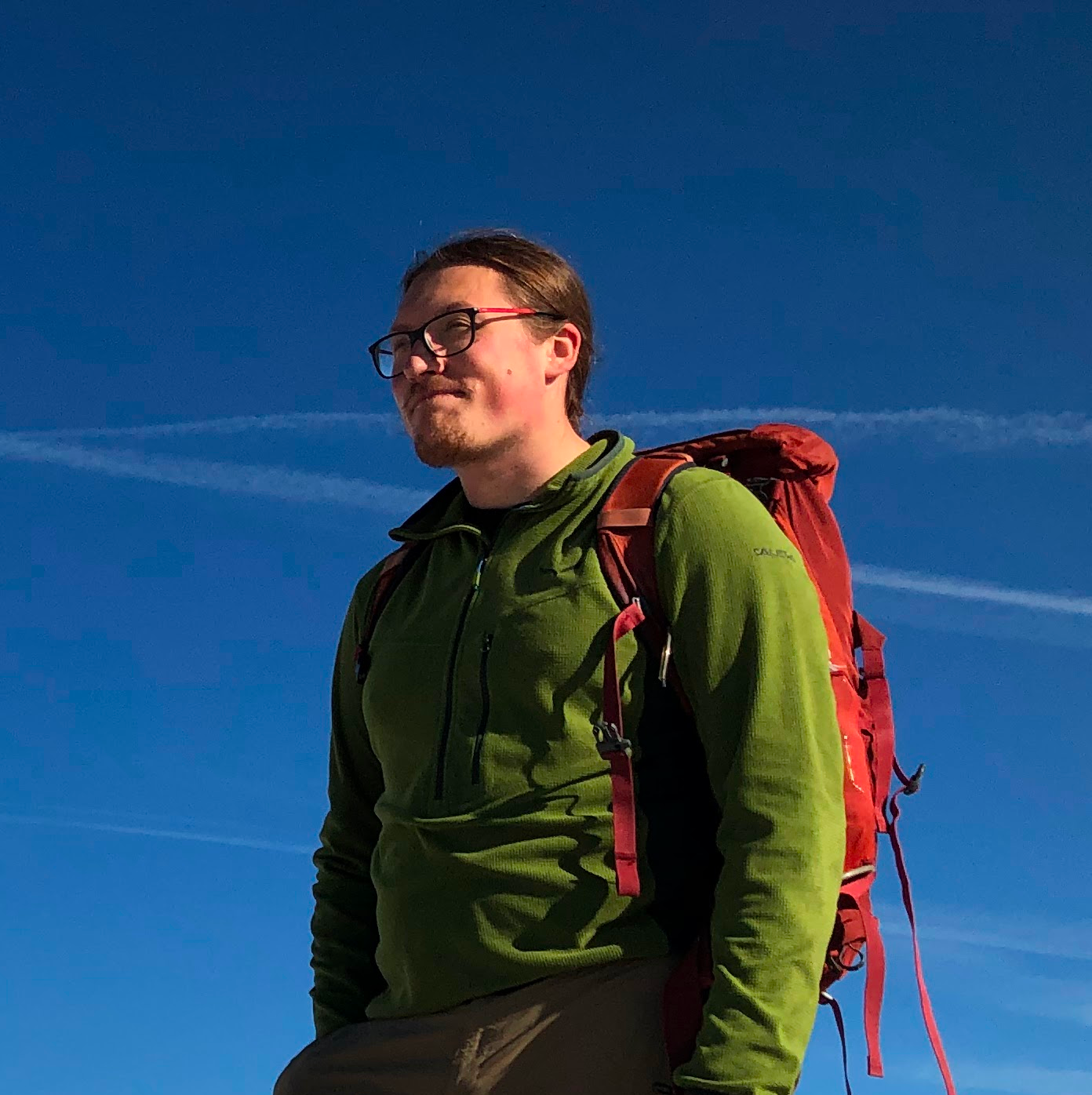}}

\begin{document}
\vspace*{4cm}
\title{Using gravitational wave early warning to pre-point neutron star mergers with Imaging Atmospheric Cherenkov telescopes}

\author{ Jacopo Tissino }

\address{Gran Sasso Science Institute (GSSI), I-67100 L'Aquila, Italy \\
Laboratori Nazionali del Gran Sasso (INFN), I-67100 Assergi, Italy}

\maketitle\abstracts{
The LIGO-Virgo-KAGRA (LVK) collaboration has recently made it possible for early warning alerts to be sent out, potentially before the end of the gravitational wave (GW) emission from a neutron star binary.
If we get such alerts in this (the fourth) or the next observing run they may arrive up to tens of seconds before the merger, which is comparable to the slewing times of the Large Size Telescopes (designed to observe very high energy gamma rays): it would be therefore possible to point to the source right before it starts emitting an electromagnetic signal.
This new mode of observation would allow us to detect the TeV component of prompt emission, which is currently poorly constrained and understood.
There are many technical challenges to overcome before this can be realized: improving the synergy between gravitational observatories and telescopes, reducing operational latencies and, from the gravitational wave side, providing more information, such as real-time updates on early warning candidates and the probability distribution of the inclination angle.
Although we may need to wait a few years---in the worst case scenario, until the next generation of GW detectors is built---before the first detection of this kind is made, implementing these improvements is a necessity.}

% \section{Multimessenger astronomy with early warning}

Neutron star binaries are a prime candidate for multimessenger astronomy: they emit gravitational waves (GWs) during their inspiral, followed by multi-wavelength electromagnetic emission.
The first coincident detection of a neutron star binary in gravitational waves with a short gamma ray burst, GRB 170817A, \cite{gw170817,grb170817a} on the 17th of August 2017, was made by Fermi-GBM and INTEGRAL.
Information about the localization of the source from the gravitational waves would only become available hours later, while the delay between merger and burst was less than 2 seconds: the discovery was made serendipitously, these wide-field instruments fortunately happened to be oriented in such a way that the emission was detectable.

While we hope to get more of this kind of detections, another higher-effort avenue has the potential to yield a better result: a pointed follow-up for energies from the keV to the TeV, as is done in the optical and other bands.
Achieving this would make it possible to use the sensitivity of instruments like Imaging Atmospheric Cherenkov Telescopes (IACTs) or the localization capabilities of X-ray facilities like Swift-XRT.
Unlike lower-energy bands, however, the detectable part of high-energy emission is often short-lived and quickly decaying: detecting it requires fast communication with the observatories.
Let us then discuss the current low-latency infrastructure.

The fourth observing run of the LVK collaboration, dubbed O4, started in spring 2023 and will continue until the end of 2025.
Since the start of O4, the possibility of sending Early Warning (EW) alerts has been implemented.\cite{llai}
The in-band duration of the gravitational chirp from a neutron star binary is typically quite long, on the scale of one to a few minutes, meaning that it can be analyzed and localized before it ends.\cite{ew1}
This task is performed employing truncated template banks, which end at various times, ranging from around 60 to around 10 seconds before merger.\cite{ew1}

As of now (the end of May 2024), only five early warning alerts have been issued, but all  have been retracted,\footnote{Information about this is public and can be obtained by querying the Gravitational wave Candidate Event Database (GraceDB: \url{gracedb.ligo.org/search/}) with the keyword \texttt{EARLY\_WARNING}.} as none of them were followed by a full-bandwidth alert; for comparison, the total number of retracted alerts was 14.
The decision to retract a lone early warning alert is much easier than the typical considerations regarding retractions: if it was not followed by a full-bandwidth alert, it is surely false, with no need to consider other factors.

The goal of early warning alerts is to communicate information about an event fast enough that actions can be taken soon, ideally before the merger.
The technical challenge this poses is great: after the gravitational waves from a binary merger reach the interferometers, several things need to happen.
(i) The strain data needs to be calibrated and distributed between computing centers, (ii) analyzed by search pipelines which will give an alert, (iii) which is then handled by the event orchestrator \texttt{gwcelery}, (iv) input into the database \texttt{GraceDB}, and (v) communicated to the larger community through a General Circular Network (GCN) or an event broker like \texttt{SCiMMA}. After this, a telescope needs to (vi) ingest the information and (vii) slew in order to detect the event.

Prior to the start of O4, a study measured the typical latencies in steps (ii) through (iv) for a set of injections, produced using realistic noise from the previous observing run (O3): the median for these was around 30 seconds after merger for all alerts, and around 3 seconds \emph{before} merger for early warning alerts; the tail of the distribution, on the other hand, reached down to around a minute before merger.\cite{llai}
Step (i) was not evaluated, and it typically takes around 10 seconds, which already shifts the median as well as the bulk of the latency distribution for early warning alerts after the merger. 

When the alert is received by a telescope optimized for it, the response can be remarkably fast: for example, the Gamma Ray Burst (GRB), GRB 160821B was detected by the wide-field Burst Alert Telescope (BAT) on board the Swift satellite, and the communication about it was received by the Major Atmospheric Cherenkov Telescopes (MAGIC) within 13 seconds (analogous to items (v) and (vi) in the previous nomenclature), with the observation starting only 11 seconds later (vii), for a total of 24 seconds of latency.\cite{magic2016}

The general picture is that, for the current observing run, having the response of a telescope start before the merger is not impossible, but it would require an extremely loud event occurring under almost perfect circumstances.
Hence, the first study\cite{ewet} proposing a very-high-energy (VHE, E$>60\,$GeV) $\gamma$-ray observation performed in response to gravitational early warning was in the context of a future instrument: the Einstein Telescope, planned for the 2030s with an extremely good sensitivity in the low-frequency band, which will make it especially suited for early warning.

This study highlighted the great scientific return that would be achieved with such an observation: the early and variable part of the emission from short GRBs, called \emph{prompt}, is believed to originate through internal dissipation within the ultrarelativistic jet, which then decelerates in the surrounding medium, giving rise to the much better understood \emph{afterglow} phase.
What is the mechanism of this deceleration? What is the composition of the jet? These questions may be answered at least in part if we had information about the VHE component of the emission, which we expect to be generated through synchrotron self-Compton (SSC).\cite{ewet,camel-boat}

\begin{figure}[ht]
    \centering
    \includegraphics[width=.8\textwidth]{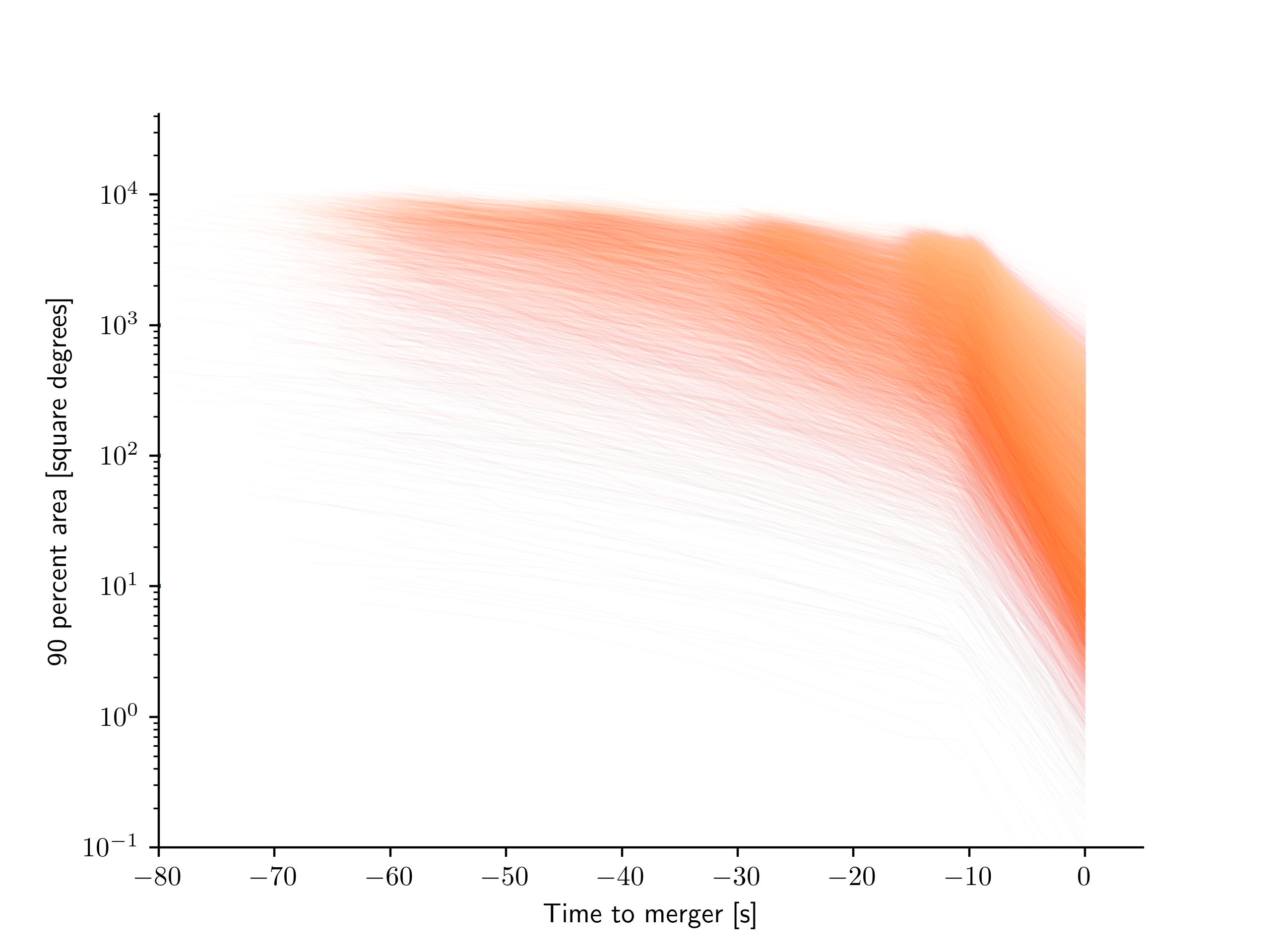}
    \caption[Evolution of the skymap area with time]{
        Evolution of the skymap area with time. Data from the injections performed by Sachdev et al.\cite{ew1}
        }
    \label{fig:skymap-area-evolution}
\end{figure}

Let us discuss another issue: the early localization of gravitational transients, which is typically performed during step (iii) using \texttt{BAYESTAR}.\cite{bayestar}
Sky localization areas for gravitational events tend to be broad, which is especially true for early warning alerts: figure \ref{fig:skymap-area-evolution} shows the evolution of the 90\% localization area with time for various events with early warning, assuming design sensitivity for the LIGO and Virgo detectors.
% I know this doesn't mean anything - I think they're referring to design O4 sensitivity, ~160-190Mpc for all three detectors.
Very few events are expected to be localized within less than 100 square degrees earlier than 10 seconds before the merger.

With this in mind, optimizing the observation strategy is crucial. Banerjee et al.\cite{ewet} identified two main modes of observation for the VHE counterpart of a GW signal, which is expected to be extremely bright, potentially requiring mere tens of seconds of exposure for a significant detection in the early stages: (a) one-shot observation, for events with a localization area small enough to be significantly covered by the field of view, and (b) mosaic tiling, which is currently employed to search for late counterparts (afterglow emission).

Only strategy (a) can allow us to detect prompt emission in the VHE band. With the sensitivity of next generation detectors we may achieve small localization areas well in advance of the merger, but as discussed earlier, this is not feasible in the next observing run (O5).
However, we propose that early alerts (potentially with some improvements compared to their current LVK implementation) can be useful even if they do not precisely localize the event for the one-shot strategy to be applicable.

For one, even a thousand-degree localization area can define a rough direction in the sky, allowing a telescope to start slewing. If several early updates were sent for the same event, the telescope could progressively refine its pointing as it receives more information, lowering the time required to have the true location in the field of view once the area becomes small enough.
Further, if a sequence of alerts were issued as updates to the first early one, some steps of the response may be fast-tracked, further reducing latency.
Even if this were to increase the false positive rate, it would constitute a minuscule commitment in terms of telescope time: for any early alert, we can know for sure whether it was a false alarm within a minute at most, and from O4 we know to expect a rate of a few per year for these. In the most pessimistic scenario, this would account for a half-hour per year of telescope time.
However, currently the low-latency GW infrastructure will send at most one early alert per event (technically, per ``superevent'', an aggregate of several alerts from different pipelines arrived within a short interval and believed to have the same source): we believe that this may become suboptimal in the next observing run(s).

How do we determine which IACTs are the most suited for this kind of observation? 
Surely a fast slew speed and a large field of view are beneficial.
Furthermore, their energy band needs to reach low enough: most IACTs peak in sensitivity around 1 TeV, while from the GRBs observed so far in this band we expect the second hump to peak in the 10-300 GeV range;\cite{agile-boat,camel-boat} although, it should be noted that these were long (i.e.\ likely originating from core collapsars), and no short GRB was detected in the VHE band yet.

% rom the GRBs observed so far with IACTs, such as GRB 190114C (ref), GRB 190829A, GRB 201216C (ref), the peak of the second hump is expected to be around sub-TeV energies. Moreover, several studies showed that the second component 
 % related to prompt emisison observed in GRB 221009A peaks between 10 GeV -300 GeV (ref to our BOAT paper, AGILE paper). Thus this makes the IACTs suitable for the GRB hunting. 

In terms of fast slew speed and energy band, the MAGIC telescope and the CTA-LST are the best candidates, slewing as fast as 6 degrees per second, and reaching less than 100\,GeV in energy.\cite{lst-magic}
The ASTRI Mini Array,\cite{astri} on the other hand, slews slower and has a higher energy threshold, but its field of view is almost three times as large as those of the aforementioned telescopes, allowing for a higher probability of catching the source with a single pointing.

We are currently exploring the effect these strategies would have in terms of increasing the number of expected detections, their practical feasibility, and the changes they will require from both the gravitational wave side and the telescope one.


\section*{References}

\begin{thebibliography}{99}

\bibitem{gw170817}Abbott et al., \Journal{\PRL}{119}{16}{2017}

\bibitem{grb170817a}Ajello et al., \Journal{\ApJ}{861}{2}{2018}

\bibitem{llai}S. S. Chaudary, A. Toivonen et al., \Journal{\PNAS}{121}{18}{2024}.

\bibitem{ew1}S. Sachdev et al., \Journal{\ApJL}{905}{2}{2020}.


\bibitem{bayestar}L. Singer, L. Price, \Journal{\PRD}{93}{2}{2016}.

\bibitem{magic2016}V. A. Acciari et al., \Journal{\ApJ}{908}{90}{2021}

\bibitem{ewet}B. Banerjee et al., \Journal{\AandA}{678}{A126}{2023}

\bibitem{lst-magic}H. Abe et al., \Journal{\AandA}{680}{A66}{2023}

\bibitem{astri}S. Vercellone et al., \Journal{\JHEA}{35}{1-42}{2022}

\bibitem{agile-boat}M. Tavani et al., \Journal{\ApJL}{956}{1}{2023}

\bibitem{camel-boat}B. Banerjee et al., preprint, \url{https://arxiv.org/abs/2405.15855} (2024)

\end{thebibliography}
\end{document}